\documentclass[a4paper,conference]{IEEEtran}

\usepackage{geometry}
\usepackage{graphicx}
\usepackage{amsmath}
\usepackage{amssymb}
\usepackage[compress]{cite}
\usepackage{color}
\usepackage{subfigure}

\newtheorem{rem}{Remark}
\newcommand{\figsizee}{0.42}
\newcommand{\mar}{\textcolor{black}}

\begin{document}
\IEEEoverridecommandlockouts
\title{VLC Systems with Fixed-Rate Transmissions under Statistical Queueing Constraints}
    
\author{\IEEEauthorblockN{Marwan Hammouda and J\"{u}rgen Peissig}
\IEEEauthorblockA{
Institute of Communications Technology\\
Leibniz Universit\"{a}t Hannover, Hannover, Germany \\ Email: \{marwan.hammouda and peissig\}@ikt.uni-hannover.de}
\thanks{This work was supported by the German Research Foundation (DFG) -- FeelMaTyc (329885056).}}
\maketitle

\begin{abstract}
Cross-layer analysis has been gaining an increasing attention as a powerful tool to study and assess different quality-of-service (QoS) mechanisms in wireless networks. Regarding the physical and data-link layers, in this paper we provide a cross-layer study for visible light communication (VLC) systems operating under statistical QoS constraints, which are inflicted as limits on the delay violation and buffer overflow probabilities. We assume that the VLC access point (AP) is unaware of the channel conditions, thus the AP sends  the data at a fixed rate. Under this assumption, and considering an ON-OFF data source, we employ the maximum average data arrival rate at the AP buffer and the non-asymptotic bounds on  buffering delay as the main performance measures. Through numerical results, we illustrate the impacts of different physical and data-link parameters on the system performance.
\end{abstract}


\section{Introduction}
Recently, visible light communication (VLC) has emerged as a potential transmission technology, thanks to the remarkable advances in white light emitting diodes (LEDs) technology that enabled utilizing the visible light spectrum for data transmission along with illumination. In addition to providing the bandwidth required to meet the increasing demand on wireless services,
using LEDs for data transmission has many advantages over the radio frequency (RF) technology \cite{elgala2011indoor}. For example, LEDs are cheap, energy efficient, and installed almost everywhere in indoor scenarios for lighting. 

Most of the existing literature studies in VLC, see e.g., \cite{hammouda2017design,hammouda2018resource,armstrong2009ofdm
,dimitrov2012signal} and references therein, focused mainly on the physical layer aspects of the VLC systems. However, the increasing demand on delay-sensitive applications, as reported by Cisco\cite{Cisco}, requires involving additional constraints on the buffer dynamics at the data-link layer. \mar{In this regard, the delay performance of hybrid RF/VLC networks has been explored in few studies~\cite{rahaim2011hybrid,shao2016delay}. For instance, the authors in~\cite{shao2016delay} described the system delay in hybrid RF/VLC systems with and without bandwidth aggregation, while each queue in the system is modeled as an M/M/1 queue. Different than these studies, in this paper we follow another approach and we provide a cross-layer study by considering statistical quality of service constraints on the buffer dynamics. The new approach mainly benefits from the large deviation techniques such that we can guarantee that the buffer overflow (or delay violation) probabilities decay exponentially
with the target bound. One of the fundamental advantages of this approach is that closed-form expressions can be obtained, thus deeper insights on the system performance can be obtained. In this line of research, cross-layer analyses} regarding the physical and data-link layers were addressed by many researchers in the RF literature, see e.g., \cite{chang1994stability,hammouda2016effective,tang2007cross, ozmen2016wireless} and references therein. On the other hand, to the best of our knowledge, there are only few studies that recently investigated cross-layer concepts in VLC systems\cite{hammouda2017link,jin2016resource,jin2015resource}. 

Nevertheless, these studies are based on the assumption that the VLC access point (AP) has a full and instant knowledge about the channel conditions. Indeed, this is not a practical assumption in many indoor scenarios, e.g., as in airports and exhibition halls, where the AP is expected to serve a large number of users, thus learning the channel of each user can be difficult and a resource-consuming task. Furthermore, notice that the authors in\cite{jin2016resource,jin2015resource} focused on the case of constant data arrival rates at the transmitter buffer, which may not be realistic in many practical settings.

In this paper, we focus on a VLC system that operates under statistical QoS constraints, which are applied as limits on the buffer overflow and delay violation probabilities, and considering an ON-OFF data source. Different than the aforementioned cross-layer studies in VLC, we assume that the VLC AP has no knowledge about the user channel gain, thus the AP sends the data with a fixed rate.   
For such a system, we  provide a cross-layer study regarding the physical and data-link layers by employing the maximum average arrival rates at the transmitter buffer and the non-asymptotic bounds on buffering delay as the main performance metrics. \mar{To summarize, the main contributions of this paper are \textit{i}) VLC systems with a practical assumption such that no knowledge about the channel  quality is required at the access points, \textit{ii}) cross-layer analyses when the system is operating under statistical quality of services constraints, imposed as limits on the buffer dynamics, and considering an ON-OFF data source, and \textit{ii}) non-asymptotic bounds on the buffering delay. We emphasize that, to the best of our knowledge, the analytical framework provided in this paper has not been addressed by other studies yet.}

\section{System Model}
\label{sec:System_Model}

We target a point-to-point VLC network in which one LED-based access point (AP)\footnote{In this paper, we use the terms "AP" and "transmitter" interchangeably.} provides communication services to one user within the indoor environment. Throughout the paper, it will become clear that the analytical framework provided here can be easily extended to a more general multi-AP and multi-user scenario. We consider a downlink scenario  and assume that the AP is equipped with a data buffer. Particularly, the data initially arrives at the AP from a source (or sources) and is stored in the data buffer before being divided into packets and transmitted to the user in frames of $T$ seconds. Throughout this paper, we assume that the user is randomly located within the AP coverage area.  

\begin{figure}
\centering
\includegraphics[width=0.35\textwidth]{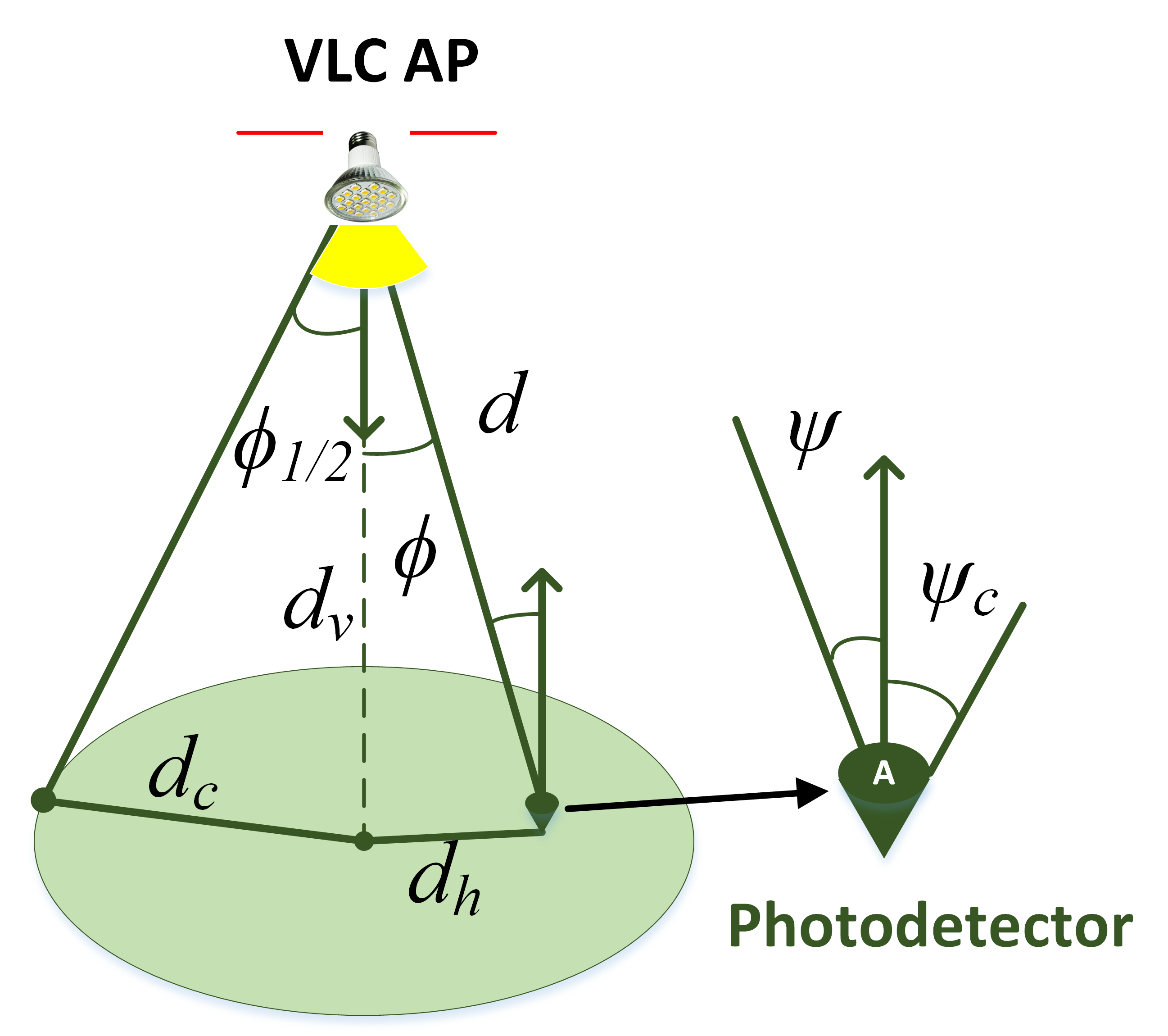}
\caption{VLC channel via LoS link.}
\label{VLC_LOS}
\vspace{-0.3cm}
\end{figure} 

\subsection{VLC Channel Model}
VLC channels normally contain both  line-of-sight (LoS) and non-LOS parts. Nevertheless, it was observed in ~\cite{komine2004fundamental} that, in typical indoor environments the main received energy at the photodetector (more than $95 \%$) comes from the LoS component. Subsequently, and without any loss of generality in this paper we only consider the LOS path, as  depicted in \figurename~\ref{VLC_LOS}. We further assume that the LED-based AP follows the Lambertian radiation pattern, and that the VLC AP is directed downwards and the user photodetector (PD) is directed upwards. Then, the LOS channel gain at a horizontal distance $d_h$ from the cell center is expressed as~\cite{barry1993simulation}
\begin{equation}
\label{eq:VLC_gain}
h = \frac{(m+1)A L(\psi) g(\psi) d_v^{m+1}}{2 \pi (d_v^2 + d_h^2)^{\frac{m+3}{2}}} \text{rect}(\psi/\psi_C).
\end{equation}
where $d_v$, $A$, and $\psi$ are, respectively,  the vertical distance, the PD physical area and the angle of incidence with respect to the normal axis to the receiver plane. In addition, $L(\psi)$ is the gain of the optical filter and $g(\psi) \frac{n^2}{\sin^2(\psi_C)}$ is the optical concentrator gain at the receiver, 
where $n$ is the refractive index and $\psi_C$ is the field of view (FOV) angle of the receiver. In (\ref{eq:VLC_gain}), $m = -1/\log_2(\cos(\phi_{1/2}))$ 
is the Lambertian index, where $\phi_{1/2}$ is the LED half intensity viewing angle. Finally, $\text{rect}(z)$ is an indicator function such that $\text{rect}(z) = 1$ if $z \leq 1$ and $\text{rect}(z) = 0$ otherwise. 

Throughout this paper, we consider the achievable rate for the VLC link of the form \cite{chaaban2017fundamental}

\begin{equation}
\label{eq:Rate_VLC}
R = \frac{T B}{2} \log_2 \bigg (1+ \frac{ (\mu \alpha P h)^2}{\varsigma^2 \sigma_{n}^2} \bigg) \quad \text{bits per frame}
\end{equation}
for some constant $\mu$. For example, setting $\mu = \sqrt{e/2 \pi}$ defines the achievable rate when the transmitted light intensity is exponentially distributed~\cite{lapidoth2009capacity}. Above, $B$ is the available bandwidth of the VLC channel and $\alpha$ is the optical-to-electrical conversion efficiency of the PD. Further, $P$ is the average transmission power limit\footnote{Notice that a peak intensity constraint can also be imposed for practical and safety concerns. However, we ignore such a limit in this paper for the sake of simplicity.}, $\sigma_n^2$  is the power of the zero-mean AWGN noise at the receiver, and $\varsigma$ is the ratio between the average optical power and the average electrical power of the transmitted signal. Setting $\varsigma = 3$ can guarantee a neglected clipping noise, thus the LED can be assumed to be operating in its linear region\cite{wang2015dynamic}.

\subsection{Fixed-Rate Transmission}
\label{sec:fixed_rate_policy}
In this paper, we assume that the VLC AP has no prior information about the channel gain $h$, or equivalently the exact user location. Thus, the  AP sends the data at the fixed rate of $\rho$ bits/frame.
 Based on the known fact that  the decoding error is negligible when transmitting at rates less than or equal to the instantaneous channel rate~\cite[Eq. (10)]{letzepis2009outage} \cite{ozarow1994information}, we assume that reliable communication is achieved when $\rho \leq R$ and thus the transmitted data is correctly decoded. On the other hand, we assume that an outage occurs when $\rho > R$, thus no reliable communication is guaranteed and data re-transmission is needed.  

For an analytical presentation, we  adopt a two-state ON-OFF Markov process to model the aforementioned VLC channel\footnote{A similar approach was also considered in many research studies to model several RF systems under different conditions and settings, see e.g., \cite{qiao2009energy,akin2010effective}. \mar{Such an ON-OFF model can also be used to describe different physical characteristics of the channel, such as LOS blockage (see e.g., \cite{wang2015efficient}) and inter-symbol interference.}}. In particular, we assume that the channel is in the ON state if $\rho \leq R$, whereas the channel is in the OFF state if $\rho > R$. Notice that a number of $\rho$ bits will be transmitted and successfully received, hence removed from the AP buffer, in one frame when the channel is in the ON state, while the transmission rate is effectively zero if the channel is on the OFF state. Here, the transition probability from the ON state to the OFF state is denoted by $\gamma_c$, and the transition from the OFF state to the ON state is denoted by $\beta_c$. Furthermore, let $p_{\text{o,c}}$ and $p_{\text{f,c}}$ be the steady-state probabilities of the channel being in the ON and OFF states, respectively, where $p_{\text{o,c}}+p_{\text{f,c}}=1$. We can easily show that\footnote{If $
	J=\left[\begin{array}{cc}
		1-\gamma & \gamma \\
		\beta & 1-\beta
		\end{array}\right]$ defines the state transition matrix for a two-state Markov model, then the steady-state probabilities can be obtained by solving the following equation: $[p_{\text{o}}\text{ }p_{\text{f}}]=[p_{\text{o}}\text{ }p_{\text{f}}]J$.}
\begin{align}
& p_{\text{o,c}} = \frac{\beta_c}{\gamma_c + \beta_c} = \text{Pr}\{\rho \leq R\}  = \text{Pr}\{d_h \leq \Delta\}  \label{eq:ON_prob_1}\\
& \hspace{-0.8cm} \text{and} \notag\\
&  p_{\text{f,c}} = \frac{\gamma_c}{\gamma_c + \beta_c}= \text{Pr}\{\rho>R\} = 1- \text{Pr}\{d_h \leq \Delta\}, \label{eq:OFF_prob_1}
\end{align}
where 
\begin{equation}
\label{eq:delta_1}
\Delta = \bigg ( \bigg( \frac{(\mu \alpha P (m+1)Ad_v^{m+1}g)^2}{(2 \pi \sigma_n \varsigma)^2(2^{2\rho/(TB)}-1)}\bigg)^{\frac{1}{m+3}} - d_v^2\bigg)^{\frac{1}{2}}
\end{equation}
Notice that the value of $\Delta$ should satisfy $0 \leq \Delta \leq d_c$. For simplicity, and due to the random user distribution, we consider a block-channel model and assume that the channel status remains the same during the frame duration of $T$ s, whereas the status changes independently from one frame to another. Formally, this yields that $\beta_c + \gamma_c = 1$ and hence we have $p_{\text{o,c}} = \beta_c \quad  \text{and} \quad p_{\text{f,c}} = \gamma_c.$

\subsection{Source Model}
\label{sec:source}
Herein, we consider a two-state discrete-time Markov process with ON and OFF states to model the data source. Notice that we can utilize the ON-OFF source to describe certain practical data arrival processes. As for instance, the ON-OFF discrete model has been widely used to describe voice sources\cite{heffes1986markov,adas1997traffic}. Nevertheless, we emphasize that other source models can be easily integrated into our framework. We consider a constant data arrival rate of $\lambda$ bits/frame in the ON state, whereas no bits arriving at the transmitter buffer in the OFF state. Herein, the transition probability from the ON state to the OFF state is denoted by $\gamma_s$ and the transition probability from the OFF state to the ON state is denoted by $\beta_s$. If $p_{\text{o,s}}$ denotes the steady-state probability of the data arrival process being in the ON state, then we have $p_{\text{o,s}}=\frac{\beta_s}{\gamma_s+\beta_s}$, and the average data arrival rate at the transmitter buffer is 
\allowdisplaybreaks
\begingroup
\begin{equation}
\label{eq:avg_rate_disc}
r_{\text{avg}} = \lambda p_{o,s} = \lambda \frac{\beta_s}{\gamma_s + \beta_s}.
\end{equation}
\endgroup
\section{System Analysis}
\label{sec:analys}
In this section, we explore the performance levels that the aforementioned system can achieve. Recall that the data is initially stored in the AP buffer before being transmitted. Thus, applying certain constraints on the buffer length is required in order to control the buffer dynamics. Let $Q$ be the steady-state buffer length and $q$ be a given threshold. Herein, we are interested in the QoS constraints regarding the buffer violation probability, i.e., $\text{Pr} \{ Q \geq q\}$, in the form of~\cite[Eq. (63)]{chang1994stability}

\begin{equation}
\label{buffer_limit}
\theta = - \lim_{q \to \infty} \frac{\log \text{Pr} \{ Q \geq q\}}{q}
\end{equation}
for a given $\theta>0$, which represents the decay rate of the tail distribution of the queue length. From~\eqref{buffer_limit}, we can approximate the buffer violation probability for a large threshold, $q_{\text{max}}$, as $\text{Pr}(Q \geq q_{max}) \approx e^{-\theta q_{max}}.$ This expression implies that, for a large threshold, the buffer violation probability decays exponentially with a rate controlled by $\theta$, which is also denoted as the QoS exponent. In particular, larger $\theta$ implies stricter QoS constraints, whereas smaller $\theta$ corresponds to looser constraints.

\subsection{Maximum Average Arrival Rate}
Considering the ON-OFF source described in Section \ref{sec:source}, we initially aim to determine the maximum average arrival rate that can be supported by the VLC channel with the fixed-rate transmission policy, as described in Section \ref{sec:fixed_rate_policy}, while satisfying the QoS requirement in (\ref{buffer_limit}). To this end, we benefit from the fundamental result shown in~\cite[Theorem 2.1]{chang1995effective}. Therein, it was proved that the constraint in (\ref{buffer_limit}) is satisfied when we have $\Lambda_{s}(\theta)=-\Lambda_{c}(-\theta)$, where $\Lambda_{s}(\theta)$, and $\Lambda_{c}(\theta)$ are, respectively, the asymptotic log-moment generating functions of the total amount of bits arriving at the AP buffer and the total amount of bits served from the transmitter. Recall that we assume a block-channel model such that the channel status changes independently from one transmission frame to another. Then, we can readily express the log-moment generating function of the service process for a given rate $\rho$ as follows\footnote{For more details about obtaining the log-moment generating functions, we refer to \cite[Example 7.2.7]{chang2000performance}.}:
\begin{equation}
\label{eq:lambda_channel}
\Lambda_c(-\theta) = \log_e  \{ p_{o,c} e^{-\theta \rho} + p_{f,c} \},
\end{equation}
where $p_{o,c}$ and $p_{f,c}$ are given in (\ref{eq:ON_prob_1}) and (\ref{eq:OFF_prob_1}), respectively. Furthermore, we have 
\allowdisplaybreaks
\begingroup
\begin{align}
&\Lambda_{s}(\theta)=\log_{\text{e}}\Bigg\{\frac{1-\beta_s+(1-\gamma_s)e^{\theta\lambda}}{2}\notag \\
& +\frac{\sqrt{\left[1-\beta_s+(1-\gamma_s)e^{\theta\lambda}\right]^2-4(1-\gamma_s-\beta_s)e^{\theta\lambda}}}{2}\Bigg\}.\label{eq:Lambda_A_discrete}
\end{align}
\endgroup
Now, using the condition $\Lambda_{s}(\theta)=-\Lambda_{c}(-\theta)$, we can formulate the maximum average arrival rate that the aforementioned VLC system can support as
\begin{equation}\label{eq:delta_discrete}
\delta(\theta) = \frac{p_{o,s}}{\theta}\log_{\text{e}}\left\{\frac{1-(1-\beta_s)D}{(1-\gamma_s)D-(1-\gamma_s-\beta_s)D^2}\right\},
\end{equation}
where $D =   p_{o,c}e^{-\theta \rho} + p_{f,c}$ and $p_{\text{o,s}}=\frac{\beta_s}{\gamma_s+\beta_s}$. 
Notice that obtaining the expressions of $\delta(\theta)$ follows from the following two main steps. First, given the log-moment generation functions in (\ref{eq:lambda_channel}) and (\ref{eq:Lambda_A_discrete}) we solve the equation  $\Lambda_{s}(\theta)=-\Lambda_{c}(-\theta)$ with respect to $\lambda$. We then obtain $\delta(\theta)$ by substitution the resulting expression of $\lambda$ in (\ref{eq:avg_rate_disc}).

\begin{rem}
\label{rem:optima:MGF}
In practice, the VLC system might be designed to support certain data transmission settings with a pre-defined QoS level, which defines the type and/or the quality of the supported service. As for example, in some places like airports and exhibition halls, users might be allowed to only download text files and/or audio files with a low quality.  In such scenarios, the transmission fixed rate can be optimized to maximize the channel performance given the target QoS needs, which we denote as $\theta_t$. It follows that, the channel log-moment generating function in (\ref{eq:lambda_channel}) can be updated as 
\begin{equation}
\label{eq:lambda_channel_1}
\begin{aligned}
\Lambda_c^{\star}(-\theta_t) & = \min_{\rho\geq0} \log_e  \{ p_{o,c}e^{-\theta_t \rho} + p_{f,c} \} \\
& =  \log_e  \{ p_{o,c}^{\star}e^{-\theta_t \rho^{\star}} + p_{f,c}^{\star} \}
\end{aligned}
\end{equation}
where $\rho^{\star}$ is the optimal fixed-transmission rate and $p_{o,c}^{\star}$ and $p_{f,c}^{\star}$ are the corresponding channel ON and OFF probabilities, respectively. Notice that  $p_{o,c}^{\star}$ and $p_{f,c}^{\star}$ also depend on the transmission rate $\rho$. Intuitively, we expect that the system designed to support a given $\theta_t$ will provide lower performance levels for any other $\theta>\theta_t$ since the system is subject to stricter QoS needs. To show this mathematically, we will consider the special case of a fixed-rate arrival process, i.e., $\beta_{s} = 1$ and $\gamma_{s} = 0$. In such a case, $\delta(\theta)$ in (\ref{eq:avg_rate_disc}) reduces to
\begin{equation}
\label{eq:EC}
\delta_{EC}(\theta) = -\frac{1}{\theta} \log_e \{ p_{o,c}^{\star} e^{-\theta \rho^{\star}} + p_{f,c}^{\star} \},
\end{equation}
which is normally referred to as the \textit{effective capacity} of the service process, and it defines the maximum \textit{constant} arrival rate that the process can support for a given QoS exponent $\theta$. It was shown in \cite[Prop. 2]{tang2007cross} that the effective capacity is a non-increasing function of $\theta$, which confirms the initial claim that the performance level either degrades or remains the same, depending on $\theta_t$ and $\rho^{\star}$, for $\theta>\theta_t$. Equivalently, designing the system for $\theta_t$ guarantees that services with QoS needs $\theta \leq \theta_t$ will be supported.
\end{rem}

\subsection{Non-asymptotic Bounds} \label{par:non_asym}
As described in (\ref{buffer_limit}), the aforementioned analysis explores the system performance in the steady-state, i.e., by assuming a large number of time frames. However, non-asymptotic bounds on the  buffer overflow and delay violation probabilities are of
interest from practical perspectives. Recall that $Q$ and $q$ define, respectively, the buffer length and a given threshold. For a given buffer overflow probability $\varepsilon$, i.e., $\text{Pr}\{\mathcal{Q} > q\} \leq \varepsilon$, ~\cite[Theorem 2]{fidler2015capacity} states that a minimal bound on the queue length can be expressed as follows: $q=\inf_{c>0}\{q_{c}+q_{a}\},$ where 
\begin{equation}\label{eq:q_r}
\begin{aligned}
	&q_{c}=-\sup_{\theta}\left\{\frac{\log_{\text{e}}\left\{-\varepsilon_{c}\left[\Lambda_{c}(-\theta)+\theta c\right]\right\}}{\theta}\right\}\\
	& \text{ for }\max\left\{0,-\frac{1}{c\varepsilon_{r}}-\frac{\Lambda_{r}(-\theta)}{c}\right\}< \theta,
	\end{aligned}
\end{equation}
and
\begin{equation}\label{eq:q_a}
\begin{aligned}
	& q_{a}=-\sup_{\theta}\left\{\frac{\log_{\text{e}}\left\{\varepsilon_{a}\left[\theta c-\sup_{t>0}\left\{\Lambda_{a}\left(\theta,t\right)\right\}\right]\right\}}{\theta}\right\}\\
	& \text{ for }0<\theta<\frac{1}{c\varepsilon_{a}}+\frac{\sup_{t>0}\left\{\Lambda_{a}\left(\theta,t\right)\right\}}{c},
	\end{aligned}
\end{equation}
Above, the buffer violation probability is $\varepsilon=\varepsilon_{r}+\varepsilon_{a}$, and $\theta$ and $c$ are free parameters. In (\ref{eq:q_a}), the time-variant log-moment generating function of the arrival process, $\Lambda_{a}\left(\theta,t\right)$, is given by \cite[Eq. (21)]{fidler2015capacity}
\begin{align}
	& \Lambda_{a}\left(\theta,t\right)= \frac{1}{t}\log_{\text{e}} \bigg\{[\begin{matrix}p_{\text{o,d}}&p_{\text{f,d}}\end{matrix}] \notag\\
	&\times  \bigg(\bigg[\begin{matrix}(1-\gamma_d)e^{\theta\lambda_d}&\gamma_d e^{\theta\lambda_d}\\\beta_d&1-\beta_d\end{matrix}\bigg]\bigg)^{(t-1)} \bigg[\begin{matrix}e^{\theta\lambda_d}&0\\0&1\end{matrix}\bigg]\bigg[\begin{matrix}1\\1\end{matrix}\bigg]\bigg\}
\end{align}

Next, we target the statistical bound regarding the queuing delay in the form $\text{Pr}\{\mathcal{D} > \tau\} \leq \varepsilon$, where $\mathcal{D}$ is the buffering delay and $\tau$ is the delay threshold. Assuming that a first-come first-served strategy is employed at the transmitter buffer, then the minimal bound on the buffering  delay can be expressed as~\cite[Theorem 1]{fidler2015capacity}
\begin{equation}\label{eq:queue_delay_bound}
\tau=\inf_{c>0}\left\{\frac{q_{c}+q_{a}}{c}\right\}
\end{equation}

\begin{table}[t]
\caption{Simulation Parameters}
\begin{center}
\begin{tabular}{l|c}\hline\hline
LED half intensity viewing angle,  $\phi_{1/2}$  & $ 60^{\circ}$ \\
PD field of view (FOV), $\psi_C$  & $ 90^{\circ}$ \\
PD physical area, $A_d$ & $ 1 \,\text{cm}^2$\\ 
Modulation bandwidth, $B$ & $40$~MHz\\ 
PD opt.-to-elect. conversion efficiency, $\alpha$ &  $0.53$~A/W\\ 
Refractive index, $n$ & $1.5$\\ 
Optical filter gain, $L(\psi)$ & $1$\\ 
Noise power spectral density, $N_0$ & $10^{-21}$~$\text{A}^2/$~Hz\\
Vertical distance, $d_v$ & $3$~m\\
Transmission frame, $T$ & $1$~ms\\
\hline\hline
\end{tabular}
\end{center}
\label{tab_1}
\end{table}%

\begin{figure}
\centering
\includegraphics[width=\figsizee\textwidth]{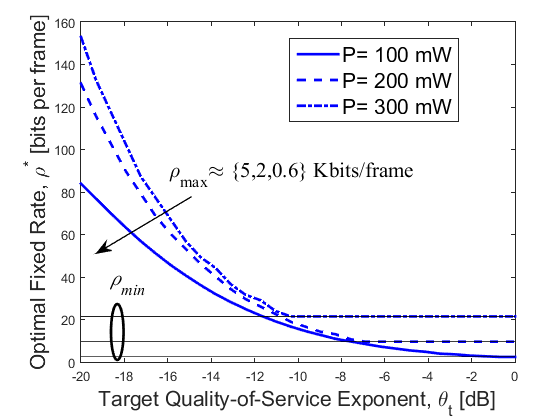}
\caption{Optimal fixed-transmission rate as a function of the target QoS exponent, $\theta_t$, and for different transmission power values.}
\label{fig:opt_rate_theta}
\vspace{-0.4cm}
\end{figure} 

\section{Numerical Results}
\label{sec:sim}
In this section, we present the numerical results. While the analytical results in the above sections are valid for any user distribution, herein we show the results assuming that the user is uniformly located within the VLC cell area. Formally, let us define the probability density function of the horizontal distance, $d_h$ as $f_{d_{h}}(h) = \frac{2 h}{d_c^2}.$
%
Subsequently, the cumulative distribution function (CDF) of $d_{h}$ can be obtained as 
\begin{equation}
\label{eq:cdf_d_h}
F_{d_{h}}(\Delta) = \text{Pr}\{d_{h} \leq \Delta \} =  \frac{\Delta^2}{d_c^2}, 
\end{equation}
where $\Delta \leq d_c$. Then, we can calculate the channel ON and OFF probabilities, i.e., $p_{\text{o,c}}$ in (\ref{eq:ON_prob_1}) and $p_{\text{f,c}}$ in (\ref{eq:OFF_prob_1}), respectively. Throughout this paper, we denote by $\rho_{\text{min}}$ and $\rho_{\text{max}}$ the minimum and the maximum achievable rate over the VLC cell, respectively, and we set the supported range of $\rho$ as $\rho_{\text{min}} \leq \rho < \rho_{\text{max}}$\footnote{Notice that we cannot set $\rho \geq \rho_{\text{max}}$ since the channel will always experience an outage, unless the user is located at the cell center and when $\rho = \rho_{\text{max}}$.}. Obviously, $\rho_{\text{min}}$ is the rate achieved when the user is located at the cell edge, while $\rho_{\text{max}}$ is the achievable rate when the user is located at the cell center. Finally, the thermal noise power at the photo-diode is $\sigma_n^2 = N_0 B$, where $N_0$ is the noise power spectral density. Table \ref{tab_1} summarizes the parameters used in this section.

In Fig. \ref{fig:opt_rate_theta}, we initially display the optimal fixed-transmission rate, following (\ref{eq:lambda_channel_1}), as a function of the target QoS needs, $\theta_t$, and for different transmission power levels, $P$. We clearly see that the optimal rate rapidly decreases with increasing $\theta_t$ to a certain value of $\theta_t$, after which the optimal rate saturates at a level that equals to the minimum achievable rate, i.e., $\rho^{\star} = \rho_{\text{min}}$. This behavior can be explained since the system randomness, either in the source or in the channel, is a key factor that limits the system ability to satisfy strict QoS requirement. Thus, stricter QoS needs can be satisfied only by reducing the randomness in the channel activity. In the VLC scenario considered in this paper, the  channel randomness can be eliminated by setting $\rho^{\star} = \rho_{\text{min}}$, since the channel stays in the ON state regardless of the user location. We further observe that increasing the transmission power results in higher possible fixed rates. To better understand Fig. \ref{fig:opt_rate_theta}, we remark the following. Noting that increasing the transmission rate, generally, results in a higher outage probability, thus reducing the system ability to satisfy stricter QoS needs, Fig. \ref{fig:opt_rate_theta} reveals that increasing the transmission power allows transmission at higher rates, and yet satisfying the QoS needs. However, the allowed rates are still far lower than the maximum achievable rate over the cell, i.e., $\rho_{\text{max}}$.  Notice that the influence  of the transmission power on the system performance has a practical significance since the power can be  related to the number of served users in multi-user scenarios, in which the transmission resources are divided among the users. Also, the transmission power shows the impact of light dimming. 
\begin{figure}
\centering
\includegraphics[width=\figsizee\textwidth]{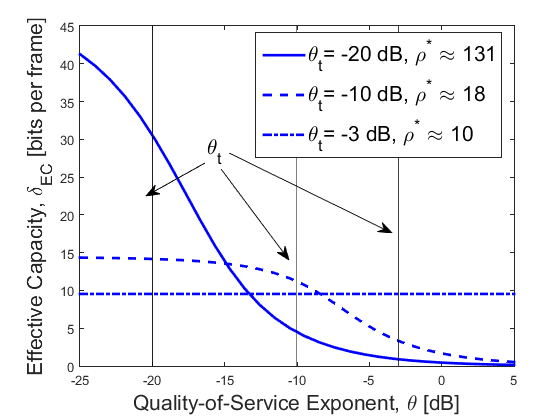}
\caption{Effective capacity as a function of $\theta$ and for different target QoS needs, $\theta_t$. Here, $P = 200$ mW.}
\label{fig:EC_theta_t_theta}
\vspace{-0.4cm}
\end{figure} 

To explore the impact of desgining the system to support given QoS needs, in Fig. \ref{fig:EC_theta_t_theta} we plot the effective capacity, given in (\ref{eq:EC}), as a function of the QoS exponent, $\theta$, and for different values of the target exponent $\theta_t$. We immediately observe that the performance curves either degrade  or remain the same with increasing $\theta$, which confirm the conclusion explained in Remark \ref{rem:optima:MGF}. Specifically, we observe that the performance gain for $\theta<\theta_t$ decreases  for larger values of $\theta_t$. This is since the optimal transmission rate increases by decreasing $\theta_t$, thus the exponential decay rate, controlled by the product $-\theta \rho^{\star}$, increases. As for instance, for $\theta_t = -3$ dB, the performance curve is flat with $\theta$, since the optimal transmission rate is equal to the minimum rate, i.e., $\rho^{\star} = \rho_{\text{min}}$, and then the effective capacity reduces to $\delta_{EC}(\theta) = \rho_{\text{min}}$.  This behavior also agrees with the results shown in Fig. \ref{fig:opt_rate_theta}.

\begin{figure}
\centering
\subfigure[$\phi_{1/2} = 60^{\circ}$]{\includegraphics[width=\figsizee\textwidth]{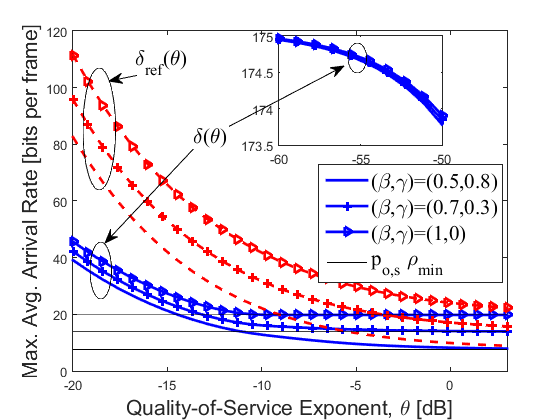}}\\
\subfigure[$\phi_{1/2} = 45^{\circ}$]{\includegraphics[width=\figsizee\textwidth]{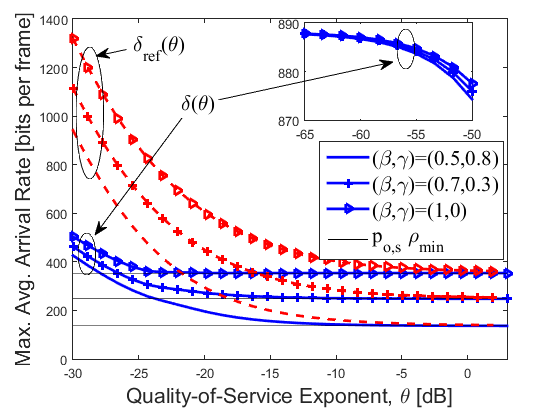}}
\caption{Maximum average arrival rate as a function of the QoS exponent, $\theta$, and for different source statistics. Here, $P = 200$ mW.}
\label{fig:compare_avg_rate_theta_phi}
\vspace{-0.4cm}
\end{figure}

In Fig. \ref{fig:compare_avg_rate_theta_phi}, we plot the maximum average arrival rate at the transmitter buffer as a function of the  QoS exponent\footnote{Int his figure, we simply set $\theta_t=\theta$.}, $\theta$, and for different source statistics. We further compare the results with those obtained with a reference scenario in which the AP has a perfect knowledge of the channel gain. Thus, the AP can send the data with a rate equals to the  channel achievable rate, i.e., $\rho = R$, where $R$ is expressed in (\ref{eq:Rate_VLC}) for a given horizontal distance $d_h$. In such a scenario, and assuming a random user distribution within the cell coverage area, the channel log-moment generating function for a give $\theta$ can be expressed as follows: 
\begin{equation}
\Lambda_{c}(-\theta) = \log_e \bigg\{\mathbb{E}_{d_h} \{e^{-\theta R}\}\bigg\},
\end{equation}
and the maximum average arrival rate, denoted as $\delta_{\text{ref}}(\theta)$, has the same expression as in (\ref{eq:delta_discrete}) with $D = \mathbb{E}_{d_h} \{e^{-\theta R}\}$.

In Fig. \ref{fig:compare_avg_rate_theta_phi}, we immediately observe that the supported average rate decreases with increasing $\theta$. We further notice that learning the channel gain has a positive impact on the system performance at lower QoS levels, whereas the performance curves of both scenarios approach the same level, for given source statistics, as $\theta \to \infty$ . This observation also means that, while the reference scenario outperforms the fixed-rate scenario at lower $\theta$ values, the reference scenario is affected more by increasing $\theta$, as the decreasing rate of the performance curves is faster with increasing $\theta$. This can be explained due to the higher level of the transmission randomness in the reference scenario. Then, we can conclude that transmitting with a fixed rate is preferred when stricter QoS needs are required since it is simpler to implement. We further notice the following two observations. First, the average arrival rate is independent of the source statistics as the QoS exponent diminishes, i.e., as $\theta \to 0$. This is  sensible since the average arrival rate is expected to approach the average transmission rate in the channel as $\theta$ goes to zero. Mathematically, we can easily prove that $\delta(\theta)$ in (\ref{eq:avg_rate_disc}) simplifies to $\delta(\theta) = p_{o,c}^{\star} \rho^{\star}$ as $\theta \to 0$. Second, the performance curves saturation at certain levels as $\theta$ increases agrees with the results shown in Fig. \ref{fig:opt_rate_theta}, that the optimal transmission rate approaches $\rho_{\text{min}}$ as $\theta \to \infty$. Subsequently, we can easily show that $\delta(\theta)$ in (\ref{eq:avg_rate_disc}) reduces to $\delta(\theta) =  p_{\text{o,s}} \rho^{\star} = p_{\text{o,s}} \rho_{\text{min}}$ as $\theta \to \infty$, where $p_{\text{o,s}} = \frac{\beta_s}{\beta_s + \gamma_s}$. From these two observations, we can conclude that the average arrival rate depends only on the service (channel) dynamics as $\theta \to 0$, whereas the average arrival rate depends mainly on the arrival (source) dynamics as $\theta \to \infty$. We finally notice that decreasing the cell area, i.e.,  by decreasing the transmission viewing angle $\phi_{1/2}$, improves the system performance as higher transmission rates can be supported and the channel randomness due to the user random location decreases. 

\begin{figure}
\centering
\includegraphics[width=\figsizee\textwidth]{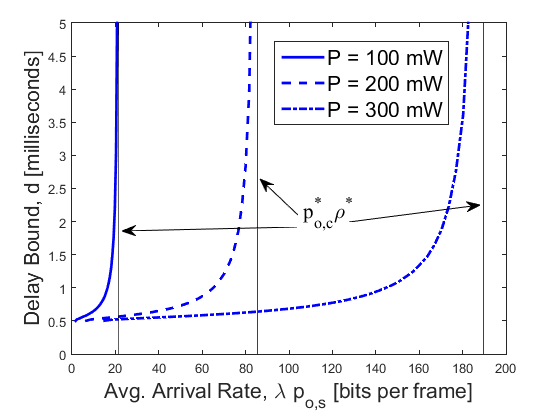}
\caption{Delay bounds as a function of the average arrival rate and considering different power levels. Here, $\gamma_s = 0.3$ and $\beta_s = 0.7$.}
\label{fig:delay_lambda}
\vspace{-0.4cm}
\end{figure} 

Finally, in Fig. \ref{fig:delay_lambda} we illustrate the delay bounds achieved by the VLC system as a function of the average arrival rates. We set $\gamma_s = 0.3$ and $\beta_s = 0.7$ and we consider different transmission power values. For each value of the transmission power, we find the value of $\theta$ that minimizes the channel-related buffer overflow probability, following (\ref{eq:q_r}). We then compute the average transmission rate that corresponds to the optimal channel log-moment generating function at that $\theta$ level, i.e., $p_{\text{o,c}}^{\star} \rho^{\star}$. The vertical lines in Fig. \ref{fig:delay_lambda} show that the delay bounds
increase asymptotically as the average arrival rate approaches
the average transmission rate in the channel, since the system becomes unstable when the average arrival rate is greater than the average service rate in the channel, and long buffering periods are expected.  We further observe the effect of increasing the transmission power on improving the delay performance.
\section{Conclusions}
\label{sec:conc}
In this paper, we have explored the performance of a VLC system with a fixed-rate transmission and an ON-OFF data source when statistical QoS constraints are applied as limits on the buffer overflow and delay violation probabilities. Regarding the physical and data-link layers, we have provided a cross-layer study by regarding the maximum average arrival rate and the non-asymptotic delay bounds as the performance metrics. Through numerical results, we have shown that transmitting with a fixed rate is preferred when stricter QoS constraints are required.  
\bibliographystyle{IEEEtran}
\bibliography{references}

\end{document}